# An Optimal ODAM-Based Broadcast Algorithm for Vehicular Ad-Hoc Networks

**Weifeng Sun[1], Feng Xia[1*], Jianhua Ma[2], Tong Fu[1], Yu Sun[3]**
[1] School of Software, Dalian University of Technology
Dalian 116620, China
[e-mail: wfsun@dlut.edu.cn]
[2] Faculty of Computer & Information Sciences, Hosei University
Tokyo 184-8584, Japan
[e-mail: jianhua@hosei.ac.jp]
[3] School of Software Engineering, University of Science and Techonology of China
Hefei 230051, China
[e-mail: sunyu123@mail.ustc.edu.cn]
*Corresponding author: Feng Xia [e-mail: f.xia@ieee.org]

## Abstract

Broadcast routing has become an important research field for vehicular ad-hoc networks (VANETs) recently. However, the packet delivery rate is generally low in existing VANET broadcast routing protocols. Therefore, the design of an appropriate broadcast protocol based on the features of VANET has become a crucial part of the development of VANET. This paper analyzes the disadvantage of existing broadcast routing protocols in VANETs, and proposes an improved algorithm (namely ODAM-C) based on the ODAM (Optimized Dissemination of Alarm Messages) protocol. The ODAM-C algorithm improves the packet delivery rate by two mechanisms based on the forwarding features of ODAM. The first distance-based mechanism reduces the possibility of packet loss by considering the angles between source nodes, forwarding nodes and receiving nodes. The second mechanism increases the redundancy of forwarding nodes to guarantee the packet success delivery ratio. We show by analysis and simulations that the proposed algorithm can improve packet delivery rate for vehicular networks compared against two widely-used existing protocols.

***Keywords:*** Vehicular ad-hoc networks; broadcast; routing; information redundancy; packet delivery rate

A preliminary version of this paper was presented at 2011 IEEE International Conference on Internet of Things (iThings'11). This version presents substantially new analytic and simulation results. This work is partially supported by Nature Science Foundation of China under grants No. 61103233, 61070181 and 60903153, NSFC-JST under grant No. 51021140004, the Fundamental Research Funds for Central Universities (DUT12JR08, DUT12JR10), Liaoning Provincial Natural Science Foundation of China under Grant No. 201202032, and DUT Graduate School (JP201006).

## 1. Introduction

Vehicular ad-hoc network (VANET), which is also called SOTS (Self-organizing Traffic Information System), is a high-speed mobile outdoor communication network [1]. The basic idea of VANET is that vehicles within a specific communication range can exchange their information of speed, location and other data obtained via GPS and sensors, and establish a mobile network automatically [2]. The range of single-hop transmission covers from a few hundred meters to a thousand meters, and each node acts as both a transceiver and a router, so multi-hop approaches are utilized to forward data to further vehicle [3]. Compared with traditional multi-hop, self-organizing networks without central nodes, there are several special features of VANET, including e.g. short path life, strong ability of computing and huge storage, high-speed mobile nodes which result in a rapid change of network topology, the ability of nodes to obtain power energy through vehicle engine, the ability of vehicle space to ensure antenna size and other additional communication equipment. In addition, nodes move in a regular pattern, mostly in single-way or two-way lane, with the feature of one-dimension, and the vehicle track is generally predictable [4].

In recent years, the relevant researches on VANET have become emphasized on a global scale. Many countries have carried out research projects on vehicle communications. For instance, the Vehicle Safety Communication (VSC) project carried out by a vehicle safety communication association consists of a number of world-renowned automobile manufacturers are also concerned about the security provided by V2V communications DSRC standards [5]. The Inter-Vehicle Network Technologies Project of New Jersey Institute of Technology focuses on the research of vehicle communications, including reliable information transmission, distributed mobile service requests, reliable routing protocols and security cooperation [6]. The Internet security group in University of Southern California analyzed the network vulnerability, and carried out a lot of work in security communications and routing protocols [7]. The Vehicular Networks Security Project of Swiss Federal Institute of Technology has been carried out by many researches to design security architecture, trust and privacy protection mechanisms in preparation for the communications within vehicles [8].

Vehicles usually communicate with base stations with WLAN or 3G. The routing technology is already relatively mature. Design of routing protocols focuses on the communication between vehicles [9]. The rapid changes of network topology caused by mobile nodes have brought a lot of difficulties in designing routing protocols. Traditional routing protocols could not adapt the high-dynamic vehicle network environment because of the delay in establishing paths [10]. Although there are some routing protocols based on self-organized network and VANET, we still need to study the nature and specific application requirements and design effective routing protocols, especially to solve the existing problem of general low packet delivery rate in VANET communications [11].

Current research suggests that VANET has different characteristics and transmission problems compared with other self-organized network [12]. First of all, VANET is an ad hoc network applied on roads, so it has the features of mobile self-organized network, such as autonomy and no fixed structure, multi-hop routing, dynamic change of network topology, limited network capacity and better scalability. However, special conditions, such as narrow roads, environments of high-density nodes and high-speed mobile nodes, will affect the information transmission ability of VANET directly, resulting in the increase of packet loss and delay [13]. Simulation results show that in VANET, the success rate of packet transmission using traditional unicast transport layer protocol (TCP, UDP) cannot exceed 50%, and the transmission will lead to higher delay time and jitter. As a consequence, a critical issue in VANET development is to design a rational and effective routing protocol according to the characteristics of VANET.

This paper deals with broadcast in VANET. This paper analyzes the shortcomings of existing broadcast protocols for VANET, and proposes an optimal broadcast algorithm called ODAM-C based on the ODAM (Optimized Dissemination of Alarm Message) protocol. The contributions of this paper include:

1) We examine the problem that ODAM cannot transmit successfully in some circumstances, and we tackle this problem by judging the angles between sender and relay nodes to select relay node flexibly.

2) Conventional broadcast protocols in VANET may cause broadcast storm and fail to guarantee data transmission reliability. We address this problem by proposing to increase the redundancy of link and expand the storage ability of nodes. In this way ODAM-C could minimize the chance of broadcast storm and link interference within communications, and increase the packet delivery rate.

3) We conduct simulation experiments and analyze the performance of ODAM-C in three scenarios of sparse, intensive and moderate node density respectively. Simulation results show that ODMA-C could achieve a better performance than other two popular protocols in vehicular environments.

The rest of this paper is structured as follows. Section 2 reviewed some broadcast protocols in VANET. Section 3 presents the ODAM-C algorithm, including its operating principles and design. In Section 4 we conduct performance evaluation via extensive simulations. Section 5 concludes the paper.

## 2. Related Work

Due to the special features of VANET, it is a critical issue to find a reliable, stable information transmission protocol in this context [14][15]. The broadcast methods are wildly used in information transmission in VANET. Among the traffic safety applications, most applications rely on multi-hop broadcast to transmit safe information. In order to notify the relevant vehicles in time, the application will broadcast security information frequently in a short time to prevent the failure of receiving packets caused by e.g. collisions.

In the intensive traffic environments, if we don't set any restriction on the forwarding of broadcast messages, then a large number of duplicate packets will emerge in the entire network, resulting in significant increase of network load. Because the wireless channel is a shared medium, and there is no RTS/CTS mechanism in wireless communication broadcast to avoid the collision, the increase of network load will cause the increase of packet collisions and network congestion. If vehicles are in large densities and they could not determine the repetitive packet in the process of forwarding, it will cause broadcast storms [16][17], resulting the network paralysis.

Wisitpongphan et al [16] describe a scenario of 4-lane highway, in the case of relatively high density of vehicles. A safe information broadcast packet will cause broadcast storm. The average packet loss rate will reach 60% or higher. This is because almost all vehicles will receive and broadcast the message. They proposed three light weight probabilistic flooding mechanisms to inhibition the flooding, including broadcast mechanism based on weight, broadcast mechanism based on time-l, and broadcast mechanism based on time-p.

Durresi et al [18] propose a hierarchical broadcast protocol BROADCOMM designed for highway network. In BROADCOMM, highway is divided into virtual cell, and with vehicles moving. Nodes on the highway are organized into two layers: the first layer consists of all nodes in a cell; the second layer consists of the reflectors placed close to the central location of honeycomb cell. Reflectors will handle the emergency messages at a certain time interval from the same cell or an adjacent cell as a Base Station. In addition, cellular reflector will serve the emergency messages for a neighbour as a relay node, and determine the next hop. This protocol is better than others based on broadcast in message delay and routing overhead, but it can only work on simple highway network.

V-TRADE (Vector-based Tracking Detection) and HV-TRADE (History-enhanced V-TRADE) [19] are both broadcast protocol based on GPS message. They will divide the neighbour nodes into different forwarding groups according to the location and motion information. For each group, only parts of vehicles (usually called border vehicles) were chosen to continue to broadcast messages. Because the new protocol only chooses a small number of vehicles to broadcast news, it will improve the bandwidth utilization with a sacrifice of small packet loss. However, as long as there is forwarding nodes, routing overhead still exists in each hop.

EDB (Efficient Directional Broadcast) [20] and LWPP (Light Weight p-Persistence) [21] both solve the broadcast storm problems by selecting a relay node to forward the packet. In EDB, all nodes will wait for a certain time after receiving the message, and then the furthest node waits for a shortest time to send an ACK message to the source node, and forward the message. The ACK message is not only used to confirm the packet transfer, but also used to suppress the same packet forwarding of other nodes. However in LWPP, each node will calculate the corresponding transmission slot according to its distance from the source node. The node will compete based on their time slots, and only one node will eventually manage to forward the packet, while the other nodes will cancel the forwarding. However, neither EDB nor LWPP could ensure that message could reach a high packet delivery rate.

Korkmaz et al [22] and Djedid et al [23] propose two kinds of broadcast protocols, MCDS (Minimum Connected Dominating Set) and Cross-layer protocol. They both use backbone to transfer broadcast packets, in which only backbone members are responsible for forwarding, and unicast is used between backbone members. As the farthest nodes will be selected as the next hop, this situation is likely to cause transmission interruption because of the weak signal relatively.

Valery et al [24] propose PGB (Preferred Group Broadcasting) protocol, and test the performance with routing protocols. PGB mainly reduces the network load through the elimination of redundant transmission, and achieves a stable routing path. In the process of finding routing path based on broadcast, choosing the closest next hop node will increase the transmission hops, while choosing the farther node as the next hop will result in the instability of link due to the movement of vehicles or signal interference. PGB chooses node with a moderate distance as the next hop, in order to provide a stable relay node for routing algorithms. In addition, PGB decides whether to forward the broadcast message based on receivers, without the extra control information. Although PGB has outstanding performance on routing protocols, it has certain delay when broadcasting traffic information.

The methods in [25] and [26] send emergency messages to avoid collision between vehicles by means of GPS devices. Valery et al [25] propose a broadcast protocol to avoid the collision based on DSRC wireless network, which use implicit response strategy. The results show that if we could ensure distance within one second interval between vehicles, 90% of chain collision could be avoided with this protocol. Yang et al [26] propose an approach to identify the application demand.

Vasco et al [27] propose a multiple-copy geographic routing protocol for vehicular delay-tolerant networks (VDTNs), called GeoSpray. VDTN is characterized by the lack of an end-to-end contemporaneous multi-hop path, which is caused by a highly dynamic network topology, sporadic and intermittent contacts, and network partitioning due to low node density and large distances.

A straightforward solution to designing a broadcast scheme which will facilitate the fast and reliable transmission of messages to the approaching vehicles is flooding, which is a simple broadcast strategy, easy to implement. As a typical representation of broadcast, each neighbours of the sender will forward the received packets [28]. It has been proved that this mechanism achieves a good information delivery rate even among high-speed mobile nodes. However, it contains a fatal defect that it will lead to the broadcast storm, resulting in information retransmission and channel congestion, which would reduce the packet success delivery rate and increase the end-to-end delay. WPBM (Weighed Probability Based Method) protocol [29] is an improved algorithm based on simple flooding, by adding a probabilistic model to reduce the redundancy of VANET. This method is similar with flooding, and it forwards the packets with the pre-calculated probability. When the probability is one, WPBA degenerates to flooding. ODAM [30] also draws extensive attention of many research organizations. The main purpose of ODAM is to reduce the number of relay nodes and reduce network redundancy, so that the utilization of network resources will be elevated.

Concerned about the broadcast storm problem, many researchers address three main areas: 1) to select the appropriate relay nodes to forward; 2) to adjust transmission signal power according to current network situation, and 3) to ensure the packet delivery rate through implicit recognition mechanism. Lots of solutions emerged based on the above three ideas, and a large part of those solutions focus on selecting the appropriate relay nodes. The accuracy of choosing appropriate relay nodes could reduce the link redundancy and delay effectively, and avoid the broadcast storm. As a result, this paper also uses this strategy to improve the packet delivery rate.

Generally speaking, methods of selecting the appropriate relay nodes can be divided into the following categories: distance-based, cluster-based, interests-based, and channel sensing-based. Distance-based relay node selection method requires certain nodes in network to perform as relay nodes to forward. The message of relay node is associated with the distance from the source node. The distance can usually be calculated with the location information of vehicle GPS. This method reduces the network congestion and end-to-end delay effectively. Representative protocols of this kind include SMART, UMB [31], ODAM, and TRADE [32]. Cluster-based relay node selection method requires the establishment and management of a cluster network. The cluster here does not refer to the establishment of cluster in reality, but virtual clusters. Representative protocols of this kind include BROADCOMM [33] and BACKBONE [34]. Interests-based relay node selection method calculates the forwarding priority through vehicles and links information. A typical example is CARISMA [35]. Channel sensing-based relay node selection methods are often proposed based on real wireless channels, e.g. REAR [36].

These methods reduce the link redundancy effectively and inhibit the broadcast storm in the process of packet delivery. However, these methods also reduce the packet delivery rate while reducing redundancy, leading to the reduction of reliability of data transmission. Hence, we propose the ODAM-C (Optimized Dissemination of Alarm Messages Capacitor) algorithm to balance the link redundancy and the packet delivery ratio. This paper extends our previous work [37] by including substantially new results, and the method is described and evaluated in more detail. The increased redundancy of ODAM-C will not result in the significant increase in the number of broadcast packets, because we use the method of expanding the packet storage capacity of nodes. As a result, it will not increase the possibility of broadcast storm or link interference.

To solve the problem of broadcast storm, we set two information lists to record the packets received in our ODAM-C algorithm. We also set a countdown timer to control the forwarding of packets. Packets received for twice in a short time period will be recorded in the second list. Packets received many times in a short time will be dropped directly from the lists. Through this method we sacrifice a small amount of redundancy in transmission. However, we reduce the probability of packet collision and broadcast storm in the process of transmission and achieve a significant improvement in packet delivery rate.

The second problem with many existing protocols is the packet loss caused by inhibition of the same direction of forwarding. To solve this problem, we assume each vehicle is equipped with a GPS and could acquire a coordinates of relative positions of neighbouring vehicles. In ODAM-C, a node in topology will calculate the angle between itself, sender and other adjacent receivers. If the neighbour relay node is in the same direction with the destination node, then packets sent from the neighbour will be ignored; otherwise forwarding of this node will be inhibited.

## 3. Proposed Broadcast Algorithm

### 3.1 Interference Node Problem

**Fig. 1** describes a simple scenario of interference nodes. There are four nodes, A, B, C and D which represent vehicles, travelling in the same direction. Node A is the initial sending node, node B and node C are in A's transmission range. The distance between B and A is larger than that between C and A, so the waiting time of node B is shorter than that of node C. In accordance with the description of ODAM, the forwarding message of node B will be listened by node C, suppressing the forwarding of node C, which eventually results in the transmission interruption from Node A to Node D.

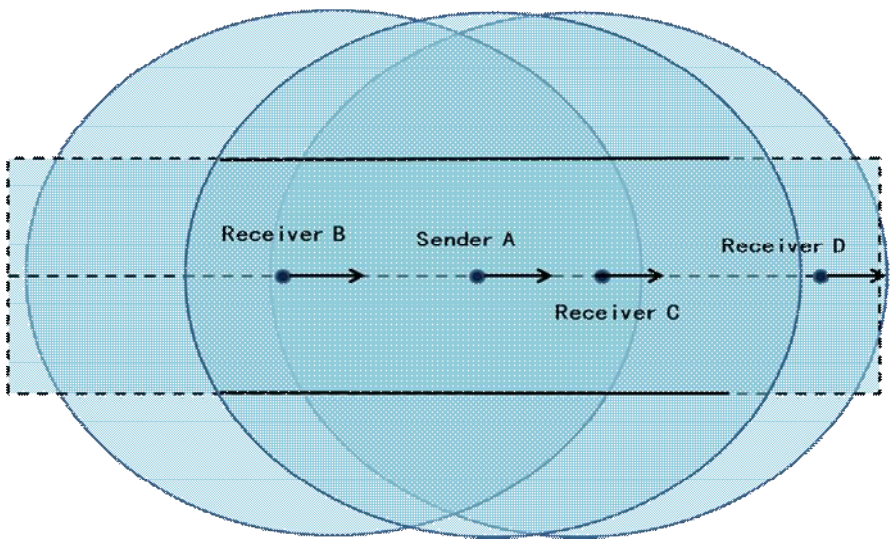

**Fig. 1.** Scenario of interference node

In ODAM algorithm, node B and node C are two relay nodes in different directions of node A, but the forward of node B inhibits the forwarding of node C, and result in sending interruption from node A to node D.

We assume each vehicle is equipped with a GPS system, so it can calculate the plane coordinates according to the latitude and longitude obtained, and broadcast its relevant information periodically, including vehicle ID, location information, speed information and level of signal intensity. Therefore, we can acquire the radius vectors and velocity vectors of vehicles in the same coordinate.

In the process of node forwarding in ODAM-C algorithm, the sender and receivers exchange information to obtain the location and other information through GPS, which can be used to calculate the angle with adjacent nodes and sending nodes. In **Fig. 1** there are four nodes node A, B, C and D moving in the same direction. Mobile node B will enter the transmission range of node A, while node C stays in the transmission range of node A. However, the distance between node A and node B is further than the distance between node C and node A, according to the distribution strategy of ODAM, node B will receive the packet earlier than node A. Assuming that the coordinate of sender A is $(X_a, Y_a)$ , the coordinate of adjacent node B, C, D is $(X_b, Y_b)$, $(X_c, Y_c)$ and $(X_d, Y_d)$, and we can calculate the angle between vector AC and vector BC through (1), (2) and (3). $\vec{a}$ and $\vec{c}$ in (1) indicate the edge-vectors of angle. In **Fig. 1**, $\vec{a}$ represents $\overrightarrow{AB}$ and $\vec{c}$ represents $\overrightarrow{AC}$ , $[(X_b - X_a) \times (X_c - X_a)] + [(Y_b - Y_a) \times (Y_c - Y_a)]$ represents $\overrightarrow{AB} \cdot \overrightarrow{AC}$ , and $\sqrt{(X_b - X_a)^2 + (Y_b - Y_a)^2} \times \sqrt{(X_c - X_a)^2 + (Y_c - Y_a)^2}$ represents $|\overrightarrow{AB}||\overrightarrow{AC}|$.

$$\vec{a} \cdot \vec{c} = |\vec{a}||\vec{c}| \cos\theta (\theta \in [0, \pi]) \quad (1)$$

$$\theta = \cos^{-1} \frac{\vec{a} \cdot \vec{c}}{|\vec{a}||\vec{c}|} \quad (2)$$

$$\angle BAC = \cos^{-1} \frac{\overrightarrow{AB} \cdot \overrightarrow{AC}}{|\overrightarrow{AB}||\overrightarrow{AC}|} = \frac{[(X_b - X_a) \times (X_c - X_a)] + [(Y_b - Y_a) \times (Y_c - Y_a)]}{\sqrt{(X_b - X_a)^2 + (Y_b - Y_a)^2} \times \sqrt{(X_c - X_a)^2 + (Y_c - Y_a)^2}} \quad (3)$$

According to the problem described in **Fig. 1**, an approach to judge inhibition of forwarding angle is proposed as follows. Step 1: each node will obtain the adjacent nodes' location information through GPS data in vehicles. Step 2: ODAM-C calculates the angles between vector AC and vector AB when sending information of node B reaches C. If the angle is less than 90 degree, and the relay node B is in the same side of node C, then the information received by node B is ignored; otherwise the delay node B is in the different side of node C, and in this situation message forwarding of node C will be inhibited.

### 3.2 Signal Attenuation Problem

**Fig. 2** describes a sparse scenario in which node B is the relay node of node A, node C and node B is the forwarding node of node C. However, if node C moves in the edge of node B's transmission range, it is hard to ensure that node C stays in the transmission range of node B at the moment that node B forwards information. An error and packet loss may occur in the process of transmission because of radio signals attenuation.

Assuming the average distance between vehicles is more than 100 meters, the density of vehicles is relatively small, so the channel information exchange is relatively less. In this case, algorithm can sacrifice a small amount of redundancy of node to increase the packet delivery rate. To achieve this aim, an information list is set in ODAM-C to collect the information received, and forward according to a countdown timer.

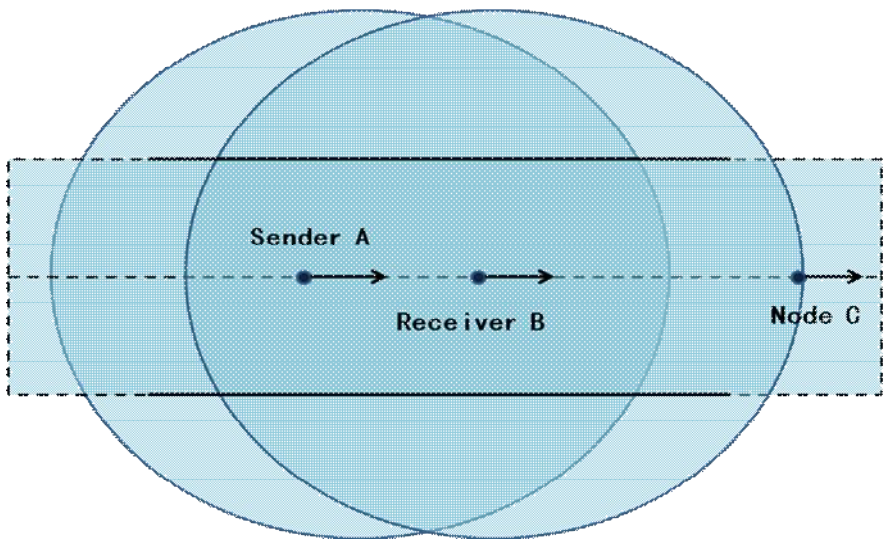

**Fig. 2.** Scenario of sparse environment

ODAM-C sets two information lists L0 and L1. Packet will be added in L1 when nodes first receive or first send packets, and forwarding timer is set as needed. After the second time the packet is received, it will be deleted from L1, added into L0, and then forwarding timer is set as needed.

**Fig. 3** describes an example of one-way forward. Here we propose another obvious example to compare with ODAM. As shown in **Fig. 3**, node A is the source sending node, node B and C are located in two different sides of A, within the transmission range of node A. When node A sends packet P6 for the first time, P6 will be added into L1 of node A, node B and node C will receive the data and add it into list. Because distance between AB is larger than AC, then node B will forward the packet first. Node A will delete P6 from L1 and insert it into L0 after receiving for the second time. As the angle of vector BCA is less than 90 degree, node C will ignore P6 and wait for the forwarding of packet P6 in L1. This case prevents the one-way inhibition of information transmission between node C and its adjacent nodes.

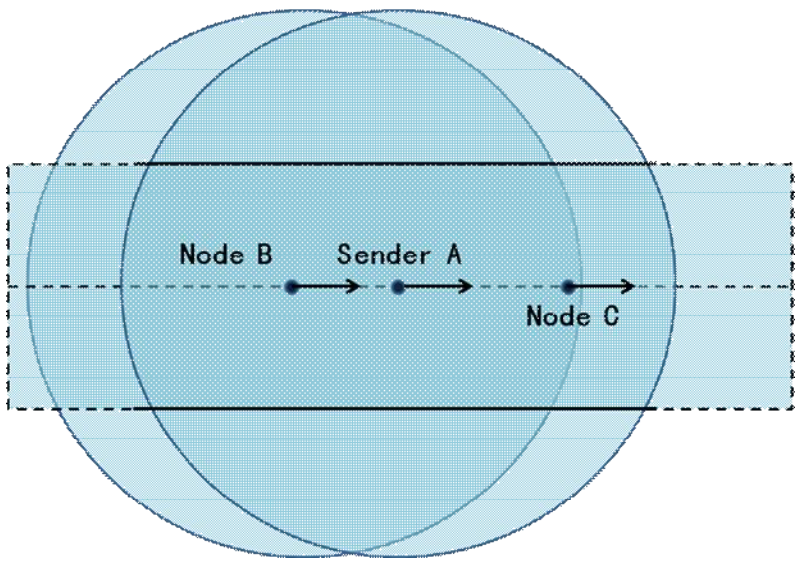

**Fig. 4.** Scenario of one-way transmission

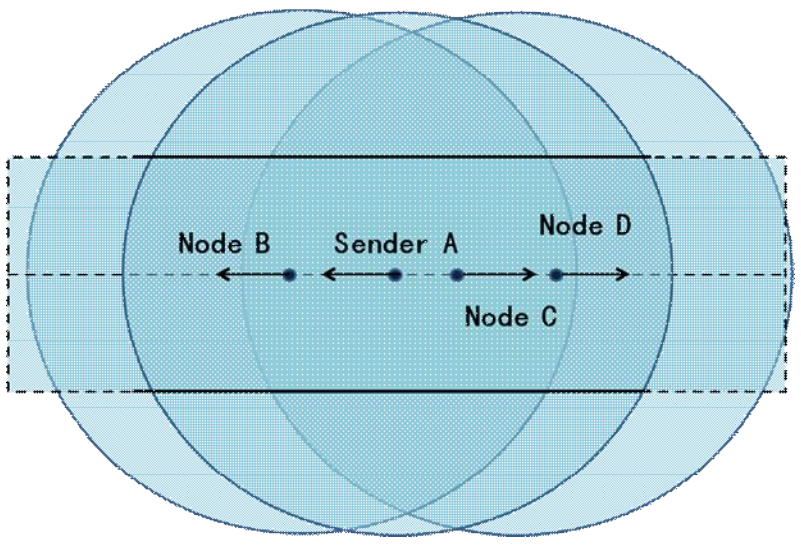

**Fig. 4.** Scenario of transmission in different flow directions

**Fig. 4** describes an example of four nodes moving in the two different directions. Node A is the sending node, node B is moving to left but node C and node D is moving to right, node B and C are in the transmission range of node A, node C is in the transmission range of node B, node A and D. After sending the packet P6, node A will add it into L1, node C, B, D receive the packet in turn and set timers after inserting P6 into L1. According to the forwarding rule of ODAM-C, node D will forward the packet first, node A and C will receive the packet from node D, and then remove the packet from L1 and add it into L0. After that node B will forward P6, node A will be suppressed after receiving P6 from B, node C notices the angle of vector DCB is larger than 90 degree (at this time node C has forwarding node on the right, and transmission of node C is inhibited, resulting in the decrease of redundancy and increase the packet delivery rate at the same time), then P6 will be forwarded in two directions of node A by node B and D.

### 3.3 ODAM-C Algorithm

ODAM-C algorithm as an improved algorithm based on ODAM is a distance-based broadcast protocol. This algorithm needs to know vehicles' location information by using GPS. The specific operation of the algorithm is based on the location data collected from the GPS or other techniques. Certain nodes will be used to forward packets in this algorithm. These nodes are selected as relay nodes. How to define relay nodes is related with their distances from source node. The relationship between waiting time and distance can be described in (4).

$$defertime(x) = max_defer_time \times \frac{(R^{\varepsilon} - D_{SX}^{\varepsilon})}{R^{\varepsilon}} \tag{4}$$

In (4), the $\varepsilon$ is a positive integer. It is used to adjust the distribution relationship between waiting time and distance, so that waiting time *defertime*(*x*) varies in the range of [0, *max_defer_time*]. The distribution is related to the value of $\varepsilon$. In the simulations the value of $\varepsilon$ is set to 2, *max_defer_time* is the twice of the average communication delay. $R$ represents the furthest transmission range of signal in a certain physical transmit frequency. $D_{SX}$ represents the distance between the sender and the receiver.

ODAM-C will store the packet received for the first time and set a timer for it. In ODAM-C, each packet is allowed to store twice at most to sacrifice a small amount of node redundancy to achieve a better packet delivery rate. ODAM-C implements this strategy by setting two packet lists. Packet lists could store useful information and timer pointers. Packets received for the first time is added to L1, and then the timer is set. The same packets received for the second time will be deleted from L1 and added in L0, and the timer is reset. More same packets will be discarded. Each packet list has a limited length. ODAM-C uses the LRU strategy to update the packet lists, ensuring the most active packet could update quickly in the list to increase the service efficiency.

ODAM-C assumes in a normal state of information exchange between vehicles in which vehicles only move in the same directions. It does not consider the case that the vehicles run in a two-way lane. The algorithm allows the overtaking and traffic jam. A description of the ODAM-C algorithm is shown in **Fig. 5**. The parameters $x$, $y$ in the algorithm represents the location information of vehicles obtained from GPS, $p$ represents the data packet, while L0 and L1 are two data lists. These are initial data input of the algorithm. In the algorithm, $d$ represents the distance between sender and receiver, *wait_time* presents the forwarding waiting time of timer calculated according to (4), and *timer_L1_p* represents the timer for packet $p$ in list L1.

Based on the algorithm description in **Fig. 5** we can know that each node makes the decision of receiving packets with packet lists L0 and L1. Each vehicle can serve as the sending node of information. For example, vehicle A is the source node to send a message for the first time. In this case, the packet will be added in L1, but node A will not set a timer. If node B receives a message for the first time, it will add the packets in L1, and set a timer for this packet.

```
Initialize x, y, p, L0 and L1
Sender:
send p
add p to L1
Receiver:
Compute d by x, y and p
Compute wait_time by d
 if p is not in L1 and p is not in L0 then
     if L1 is not  available
          update L1 with LRU algorithm
     end if
     insert p into L1
     set timer_L1_p to p
     wait for  timer_L1_p expires
            forward p
 else
     Compute angle between Receiver(x, y), Sender(x, y) and adjacent node(x, y) with Equations (1)-(3)
     if angle < 90° then
            stop forwarding after receiving p from neighbors
     else if p is in L1
            if timer_L1_p is timing then
                   stop timer_L1_p
            end if
            delete p_old from L1
            if  L0 is not available
                   update L0 with LRU algorithm
            end if
            insert p_new to L0
            set timer_L0_p to p
            wait for  timer_L0_p expires
                   forward p
     else if p is in L0 then
            if timer_L0_p is timing then
                   stop timer_L0_p
            end if
     end if
 end if
(LRU algorithm was used in ODAM-C to update the latest packet in L1 and L0)
```

**Fig. 5.** Algorithm description of ODAM-C

We assume that a lot of new data emerge at the same time in the link, caused by the problems of road or vehicles, or simply because some vehicles are sending packets constantly.

In these cases, vehicles nearby will receive new packets frequently, resulting in continually adding packets in L1. If there is no space for new packets in L1, old packets will be removed. If a packet stored in L1 is received again, packet lists will stop the timer and move the packet from L1 to L0. Packet received for the second time will be added in L0, and we will set a timer for it. This record in L0 will be the basis to calculate the redundancy. If duplicate packets are received during the countdown, then L0 stop the timer and update the packet. Packets in L1which has stopped the timer can be used to locate the redundant data of a link quickly. In this case, forwarding will be stopped to reduce the node redundancy. Packets in L0 are maintained by LRU, so they are listed in accordance with freshness, and packets will be excluded from the list if not used for a long time.

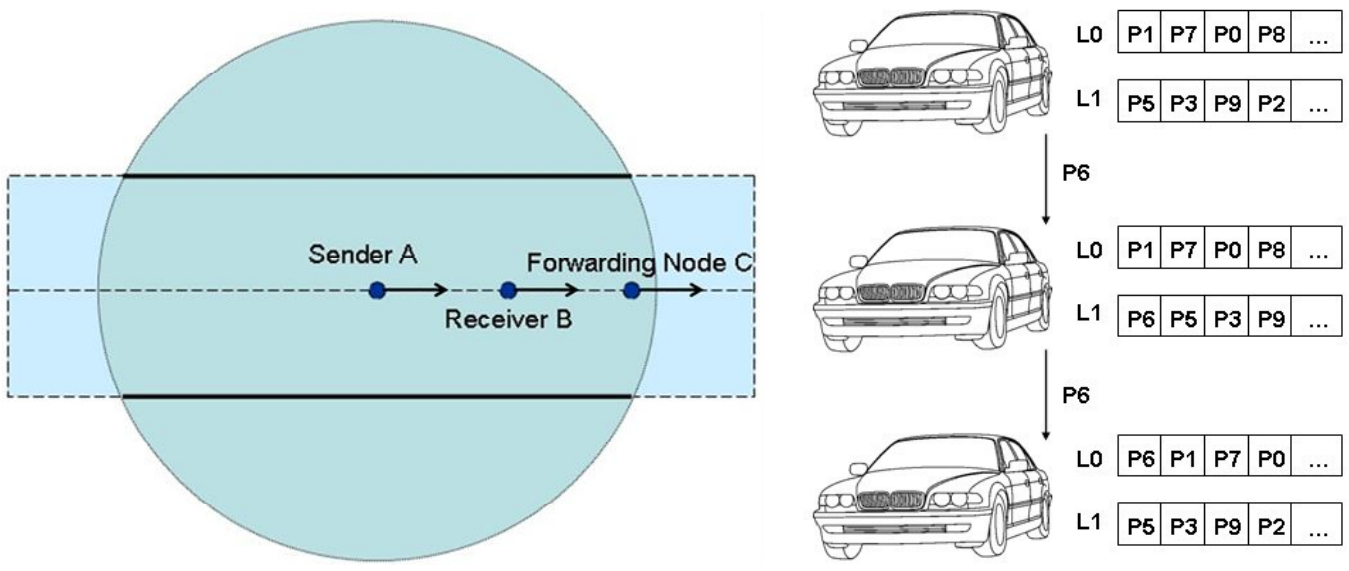

**Fig. 6.** Packet list operation

We describe a scenario illustrating the operation of packet list after receiving a packet based on the ODAM-C algorithm, as shown in **Fig. 6**. The (sub-)figure on the left is a simple transmission scenario, including sender A, receiver B and relay node C. The figure on the right shows the changing of packet list after node B receives the packet P6. In the process, other packets have not been sent out. After sender A sends a packet P6, it is received by node B for the first time. Node B adds P6 in L1, and sets a timer. Node C receives P6 and sets a timer. As node C is farther from A than B, packet P6 in node C will be forwarded before node B. After node B receives the P6 from node C, it terminates timer of P6 in L1, and removes P6 from L1, adds it in L0 and sets a timer. Since node A will forward P6 again after receiving it from node C, node A will terminate the timer of P6 in L0. Afterwards, if P6 is not removed from L1, the node B will drop the P6 directly after receiving it, so it avoids the broadcast storm problem.

# 4. Performance Evaluation

## 4.1 Performance Metrics

To evaluate the performance of ODAM-C, we conduct simulations to compare it with ODAM and WPBA. The simulations focus on three important performance metrics, which are described in Equations (5)-(7).

$$Latency = Average\ (packet\ arriving\ time - packet\ sending\ time) \quad (5)$$

*Latency* is defined as the average end-to-end latency experienced by packets that are successfully received by all destination nodes.

$$PDR = \frac{\#\ of\ received\ packets}{total\ \#\ of\ packets} \quad (6)$$

PDR (Packet Delivery Rate) is defined as the ratio of the number of successfully received packets to the total number of received packets. Higher packet delivery rate means data can be transmitted to more vehicles.

$$Redundancy = \frac{\#\, of\, forwarding\, nodes}{total\, \#\, of\, nodes} \tag{7}$$

*Redundancy* is defined as the ratio of the number of forwarding nodes to the total number of nodes. Lower redundancy leads to higher transmission efficiency and lower possibility of causing broadcast storm.

### 4.2 Simulation Settings

The popular NS-2 simulator is used to simulate a network of 200 vehicle nodes. The nodes are distributed in an area of 12000m×12000m, and work over IEEE 802.11. Signal transmission range is 300m. Channel frequency is 5.15 GHz, and transmit power is 0.281838 watt. Each node is combined with a VanetRBCAgent. VanetRBCAgent_0 begins to broadcast for the first time at 600s, and continues to broadcast every 50s. It stops broadcast at 2000s. The topography of simulation scenarios is a real-world road traffic scenario imported through SUMO (Simulation of Urban MObility) [38]. **Table 1** shows the MAC and PHY definitions for IEEE 802.11 used in simulations.

**Table 1.** MAC and PHY definitions for IEEE 802.11

| **Parameter** | **Value** |
|---|---|
| Mac/802_11 set dataRate | 6.0e6 |
| Mac/802_11 set basicRate | 6.0e6 |
| Mac/802_11 set CCATime | 0.000004 |
| Mac/802_11 set CWMax | 1023 |
| Mac/802_11 set CWMin | 15 |
| Mac/802_11 set PLCPDataRate | 6.0e6 |
| Mac/802_11 set PLCPHeaderLength | 50 |
| Mac/802_11 set PreambleLength | 16 |
| Mac/802_11 set SIFS | 0.000016 |
| Mac/802_11 set SlotTime | 0.000009 |
| Phy/WirelessPhy set RXThresh | 6.72923e-11 |
| Phy/WirelessPhy set freq | 5.15e9 |
| Phy/WirelessPhy set Pt | 0.281838 |

**Table 2** shows the configurations of simulation scenarios. In the simulations we set up three scenarios with different vehicle densities. In Scenario 1 and Scenario 2 we use 4 one-way lanes, while in Scenario 3 we use 3 one-way lanes. We configured two vehicle flows with different speeds as our mobility model. The two vehicle flows move in the same direction. High-speed vehicle has a speed of 120km/h, an acceleration of 4.5m/s$^2$, and a deceleration of 1m/s$^2$. Low-speed vehicle has a speed of 70km/h, an acceleration of 0.8m/s$^2$, and a deceleration of 4.5m/s$^2$.

**Table 2.** Simulation Configurations

| | **Scenario1** | **Scenario 2** | **Scenario 3** |
|---|---|---|---|
| Vehicle number | 100 | 200 | 500 |
| Vehicular gap (*m*) | 200 | 100 | 5 |
| Lane (*km*) | 4 | 4 | 3 |
| Length of lane (*km*) | 22 | 22 | 22 |
| Speed of Flow1 (*km/h*) | 120 | 120 | 120 |
| Acc. of Flow1 ($m/s^2$) | 4.5 | 4.5 | 4.5 |
| Dec. of Flow1 ($m/s^2$) | 1 | 1 | 1 |
| Speed of Flow 2 (*km/h*) | 70 | 70 | 70 |
| Acc. of Flow 2 ($m/s^2$) | 0.8 | 0.8 | 0.8 |
| Dec. of Flow 2 ($m/s^2$) | 4.5 | 4.5 | 4.5 |

### 4.3 Simulation Results and Analysis

**Fig. 7** shows the simulation results of redundancy in three different scenarios, indicating the data redundancy diagram of ODAM, WPBM and ODAM-C, in which *x* coordinate represents the packet ID, and *y* coordinate represents the redundancy rate of data packet.

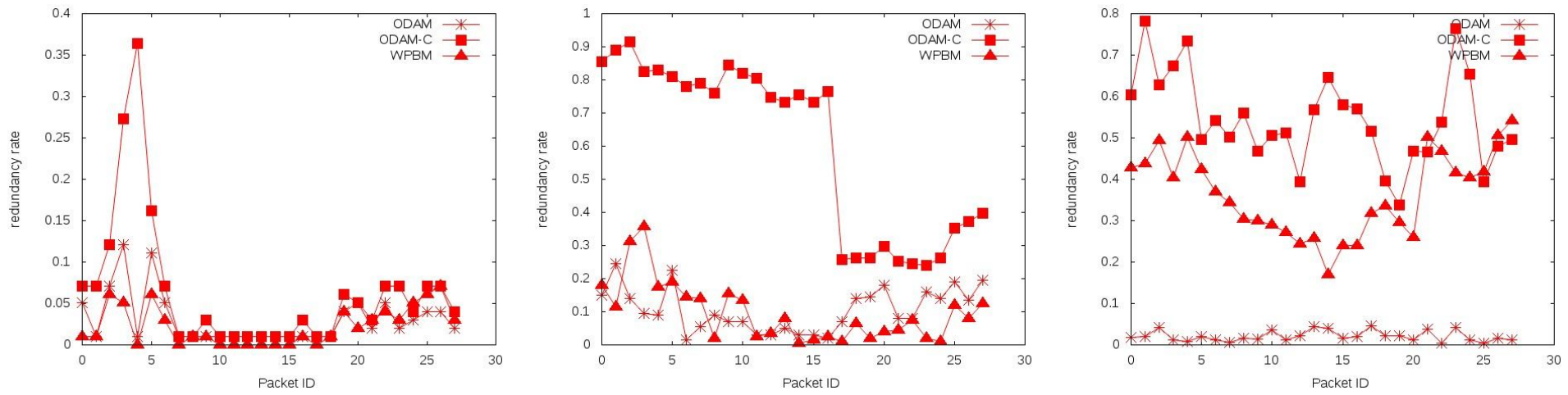

(a) Scenario1 (b) Scenario 2 (c) Scenario 3

**Fig. 7.** Results of redundancy

Three scenarios indicate the different vehicle densities of simulation environments, and the density of vehicles increases by scenario. **Fig. 7(a)** describes a scenario of low vehicle density, where the distance between vehicles is 200 meters. **Fig. 7(b)** describes a scenario of middle vehicle density, where the distance between vehicles is 100 meters. **Fig. 7(c)** describes a scenario of high vehicle density, where the distance between adjacent vehicles is 5 meters.

As we can see from **Fig. 7**, in the scenario of low vehicle density, the redundancy rate of ODAM-C changes from 0.06 to 0.36 when packet ID received is below 6; the redundancy rates of ODAM and WPBA are around 0.05; after the packet ID received reaches 6, the redundancy rates of WPBA, ODAM and ODAM-C remain almost the same (around 0.01). Except for a few cases, the redundancy of ODAM-C is quite similar with WPBM and ODAM. Generally, the redundancy of ODAM-C is slightly higher than ODAM. In the scenario of middle vehicle density, the redundancy of ODAM-C is around 0.85 when packet ID received is below 15. This is higher than those of WPBM and ODAM, which are around 0.2. The redundancy rate of ODAM-C goes down to around 0.3 when packet ID received is beyond 16. Generally, the redundancy of ODAM-C is about 3 times more than ODAM in this circumstance. In the scenario of high vehicle density, the redundancy of ODAM-C is around 0.6, which is close to that of WPBM. The redundancy rate of WPBM remains around 0.3, while the redundancy rate of ODAM is below 0.1. In general, the redundancy rate of ODAM-C is higher than that of ODAM and close to that of WPBM in high density scenario.

**Fig. 8** shows the simulation results of packet delivery rates in three scenarios, demonstrating the performance of WPBM, ODAM-C and ODAM. In the figure, the *x* coordinate represents the packet ID, and the *y* coordinate represents the packet delivery rate.

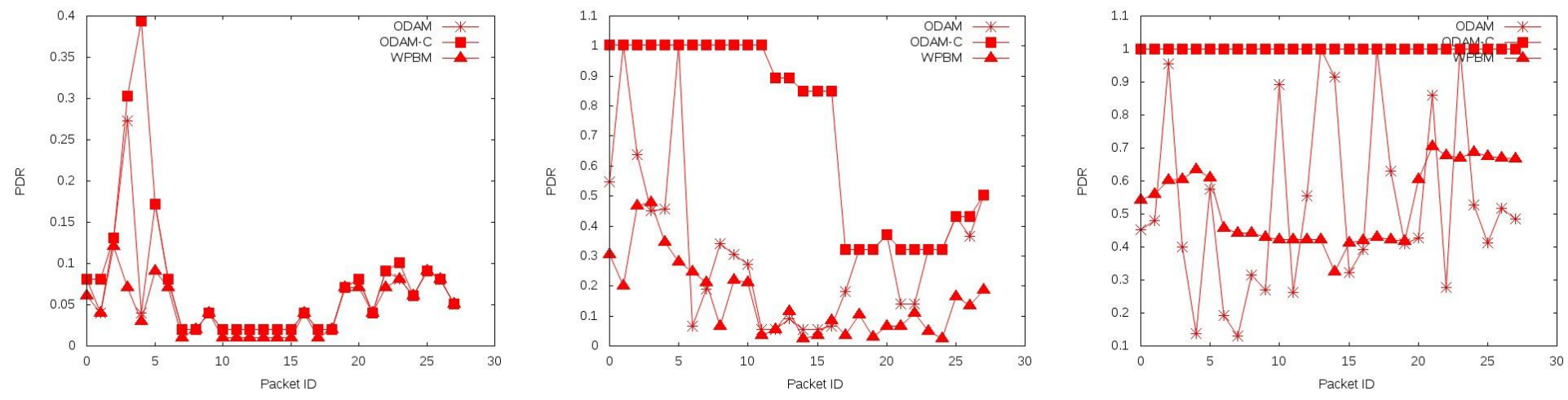

(a) Scenario1 (b) Scenario 2 (c) Scenario 3

**Fig.8.** Packet delivery rate

In the scenario of low vehicle density, when the packet ID received is less than 5, the PDR of ODAM-C ranges from 0.04 to 0.4; the PDR of ODAM ranges from 0.04 to 0.27; the PDR of WPBA ranges from 0.03 to 0.13. Except for a few cases, the PDR of ODAM-C is close to those of WPBA and ODAM. When the packet ID received is larger than 5, the PDRs of three protocols are similar.

In the scenario of middle vehicle density, the PDR of ODAM-C is close to 1 when packet ID received is less than 10, which is much higher than that of ODAM. The PDR decreases with the increase of packet number due to transmission collisions. The PDR of ODAM varies around 0.6. It is close to 1 only in two cases. The PDR of WPBA is always below 0.5. As we can see, the PDR of ODAM is slightly higher than that of WPBA, and the PDR of ODAM-C has been improved considerably as compared against ODAM.

In the scenario of high vehicle density, the PDR of ODAM-C is close to 1, much higher than that of ODAM, which varies around 0.5. The PDR of WPBA also remains around 0.55. In this set of experiments ODAM-C's PDR is the highest.

**Fig. 9** shows the simulation results of end-to-end delays in three scenarios, in which the *x* coordinate represents the packet sending time, and the *y* coordinate represents the delay of packets.

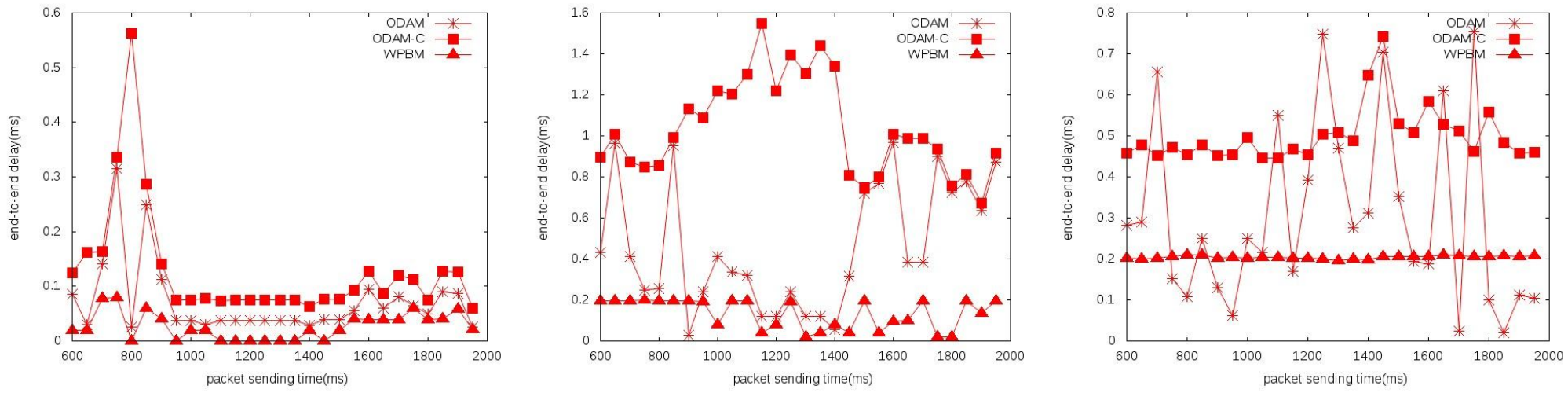

(a) Scenario1 (b) Scenario 2 (c) Scenario 3

**Fig.9.** End-to-end delay

In the scenario of low vehicle density, when the packet sending time changes from 600ms to 1000ms, the latency of ODAM-C varies from 0.12ms to 0.32ms, which is close to that of ODAM. The latency of WPBM is less than 0.1ms under this circumstance. When the packet sending time is beyond 1000ms, the latencies of three protocols are below 0.1ms. In the scenario of middle vehicle density, when the packet sending time changes from 600ms to 1000ms, the latency of ODAM-C varies around 1ms. The latency of ODAM is about 0.4ms in this circumstance. When the packet sending time varies from 1000ms to 1400ms, the latency of ODAM-C varies around 1.2ms, and it decreases when sending time reaches 1400ms. The latency of ODAM remains around 0.3ms when packet sending time is below 1400ms, and varies around 0.9ms when sending time is beyond 1400ms. The latency of WPBM is less than 0.2ms in most cases. As we can see, the latency of ODAM is slightly larger than WPBM, and in some cases the latency of ODAM-C is larger than ODAM. In the scenario of high vehicle density, the latency of ODAM-C varies around 0.5ms, which is slightly bigger than ODAM. The latency of ODAM varies from 0.1ms to 0.7ms in this circumstance. The latency of WPBM remains around 0.2ms.

From the above simulation results we can see that:

(1) In Scenario 1, the redundancy, packet delivery rate and latency of WPBM, ODAM and ODAM-C are basically the same. The redundancy and latency of ODAM-C are slightly worse, but its PDR has improved significantly.

(2) In Scenario 2, the redundancy of ODAM-C is about 3-4 times those of the other two algorithms. Its delay is about 2 times that of ODAM. The PDR of ODAM-C is about 5 times that of WPBA, and 2 times that of ODAM.

(3) In Scenario 3, the redundancy of ODAM-C is higher than ODAM, but lower than WPBA. Its latency is slightly higher than the other two algorithms, and its PDR is close to 1, much higher than ODAM.

In general, ODAM-C is on the basis of ODAM. In high-density scenario, the redundancy is slightly worse than ODAM, but the packet delivery rate increases significantly, which can reach close to 100% in some cases. In low-density scenario, the increased redundancy will also bring the growth of the packet delivery rate.

It is worth noting that the simulation results are related to the sizes of forwarding lists configured in nodes. In reality, the number of packets is relatively large, so setting an appropriate size for the list L1 could remove the old data effectively and improve the efficiency of algorithm. In the case of large variety of packets, the most active data within the transmission range of vehicles are selected and stored in data list for multiple receptions (List L0). List L0 stores data with high redundancy in the links so as to reduce the processing time of the vehicles.

## 5. Conclusion

In this paper we have analysed the shortcomings of existing broadcast protocols for VANET and proposed an optimal ODAM-based broadcast algorithm (ODAM-C). Two major problems with traditional algorithms have been addressed. The first one is the reduction of the packet loss caused by forwarding in the same direction. The other is the increase of packet delivery rate through a method of redundancy. We have implemented the ODAM-C algorithm in NS-2 and conducted extensive simulations to evaluate its performance, as compared against other two protocols, i.e. ODAM and WPBM. The results show that the proposed ODAM-C broadcast protocol provides higher packet delivery rate than the other two protocols in vehicular environments.

As future work, we will evaluate ODAM-C in a more realistic environment. Further study will focus on using vehicular traces in different scenarios (e.g. three or more vehicle flows, roads with obstacles, and vehicles moving in cross roads). In these different scenarios we will be able to evaluate extensively ODAM-C’s feasibility and scalability. In addition, algorithm cost analysis and security insurance are also important parts of our future work.